\documentclass[amsmath,amssymb,pra,aps,showpacs,twocolumn,10pt]{revtex4-1}
\usepackage{amsmath,amsfonts,amssymb,amsthm,graphics,graphicx,epsfig,bbm}
\usepackage[colorlinks=true,citecolor=blue,linkcolor=red,urlcolor=blue]{hyperref}
\usepackage[usenames]{color}
\usepackage{graphicx}
\usepackage{subfigure}
\usepackage{amsmath}
\usepackage{epsfig}
\usepackage{dcolumn}
\usepackage{bm}
\usepackage{color}
\usepackage{epstopdf}
\usepackage{amssymb}
\usepackage{amstext}
\usepackage{latexsym}
\usepackage{hyperref}
\usepackage{amsfonts}
\usepackage{psfrag}
\usepackage{xcolor}
\usepackage[normalem]{ulem}
\usepackage{dsfont}
\usepackage{txfonts}
\usepackage{overpic}
\usepackage{ifthen}

\newcommand{\ket}[1]{\vert #1 \rangle}
\newcommand{\bra}[1]{\langle #1 \vert}

\newcommand{\braket}[2]{\langle #1 \vert #2 \rangle}
\newcommand{\mean}[1]{\langle #1 \rangle}

\newcommand{\abs}[1]{| #1 |}

\newcommand{\Tr}{\mathrm{Tr}}

\newcommand{\an}[2]{\ifthenelse{\equal{#1}{}}{\ensuremath{\hat{#1}_{#2}}}{\ensuremath{\hat{#1}^{\protect\phantom{\dagger}}_{#2}}}}

\newcommand{\idg}[1]{{\bfseries #1)}}
\newcommand{\subfigimg}[3][,]{%
	\setbox1=\hbox{\includegraphics[#1]{#3}}
	\leavevmode\rlap{\usebox1}
	\rlap{\hspace*{2pt}\raisebox{\dimexpr\ht1-0.5\baselineskip}{{\bfseries \large\textsf{#2}}}}
	\phantom{\usebox1}
}

\newcommand\numberthis{\addtocounter{equation}{1}\tag{\theequation}}

\newcommand{\rev}[1]{\textcolor{blue}{#1}}

\begin{document}

\title{Topological pumping of quantum correlations}

\author{T. Haug\textsuperscript{1,2}}
\email{tobias.haug@u.nus.edu}
\author{L. Amico\textsuperscript{2,3,4,5,6}}
\author{L.-C. Kwek\textsuperscript{2,5,7,8}}
\author{W.~J.~Munro\textsuperscript{1}}
\author{V. M. Bastidas\textsuperscript{1}}
 \email{victor.bastidas@lab.ntt.co.jp}

\affiliation{\textsuperscript{1}%
NTT Basic Research Laboratories \& Research Center for Theoretical Quantum Physics,  3-1 Morinosato-Wakamiya, Atsugi, Kanagawa, 243-0198, Japan} 
\affiliation{\textsuperscript{2}%
	Centre for Quantum Technologies, National University of Singapore, 3 Science Drive 2, Singapore 117543, Singapore%
}

\affiliation{\textsuperscript{3}%
	CNR-MATIS-IMM and Dipartimento di Fisica e Astronomia, Universit\'{a} Cnia,  Via  S.  Soa  64,  95127  Catania,  Italy%
}
\affiliation{\textsuperscript{4}%
	INFN Laboratori Nazionali del Sud, Via Santa Sofia 62, I-95123 Catania, Italy %
}
\affiliation{\textsuperscript{5}%
	MajuLab, CNRS-UNS-NUS-NTU International Joint Research Unit, UMI 3654, Singapore
}
\affiliation{$^6$ LANEF \textit{'Chaire d'excellence'}, Universit\`e Grenoble-Alpes \& CNRS, F-38000 Grenoble, France}

\affiliation{\textsuperscript{7}%
	Institute of Advanced Studies, Nanyang Technological University, 60 Nanyang View, Singapore 639673, Singapore%
}

\affiliation{\textsuperscript{8}%
	National Institute of Education, Nanyang Technological University, 1 Nanyang Walk, Singapore 637616, Singapore%
}

\date{\today}

\begin{abstract}
Topological pumping and duality transformations are paradigmatic concepts in condensed matter  and statistical mechanics. In this paper, we extend the concept of topological pumping of particles to topological pumping of quantum correlations. We propose a scheme to find pumping protocols for highly-correlated states by mapping them to uncorrelated ones. We show that one way to achieve this is to use dualities, because they are non-local transformations that preserve the topological properties of the system. By using them, we demonstrate that topological pumping of kinks and cluster-like  excitations can be realized. We find that the entanglement of these highly-correlated excitations is strongly modified during the pumping process and the interactions enhance the robustness against disorder. 
Our work paves the way to explore topological pumping beyond the notion of particles and opens a new avenue to investigate the relation between correlations and topology.

\end{abstract}

\maketitle

\section{Introduction}
In 1983, D. Thouless demonstrated that the topological properties of the wave function of an extended system can be exploited to realize quantum pumps~\cite{thouless1983,Thouless1984,Seiler1985,Kwek2014,Gong2015,Gong2016,2016Tangpanitanon}. This mechanism can perform topologically-protected quantum transport, which is robust against disorder and weak interactions~\cite{Thouless1984}.
The remarkable progress of quantum technology has allowed the implementation  of Thouless's idea  to  unprecedented  degrees. 
Recently, quantum pumping of particles has been realized in diverse platforms ranging from ultracold optical superlattices~\cite{bloch2016, takahashi2016} to waveguide arrays~\cite{kraus2012,Silberberg2015}. In the context of quantum simulation, quantum pumping in low-dimensional systems can be used to simulate higher dimensional quantum systems~\cite{kitagawa2010topological}. For example, recent experiments have demonstrated that topological pumping in two-dimensional systems can be used to explore the exotic physics of Quantum Hall effect in four dimensions using cold atoms~\cite{2018Lohse} and photonic systems~\cite{2018Zilberberg}. 
 
\begin{figure}
	\includegraphics[width=0.5\textwidth]{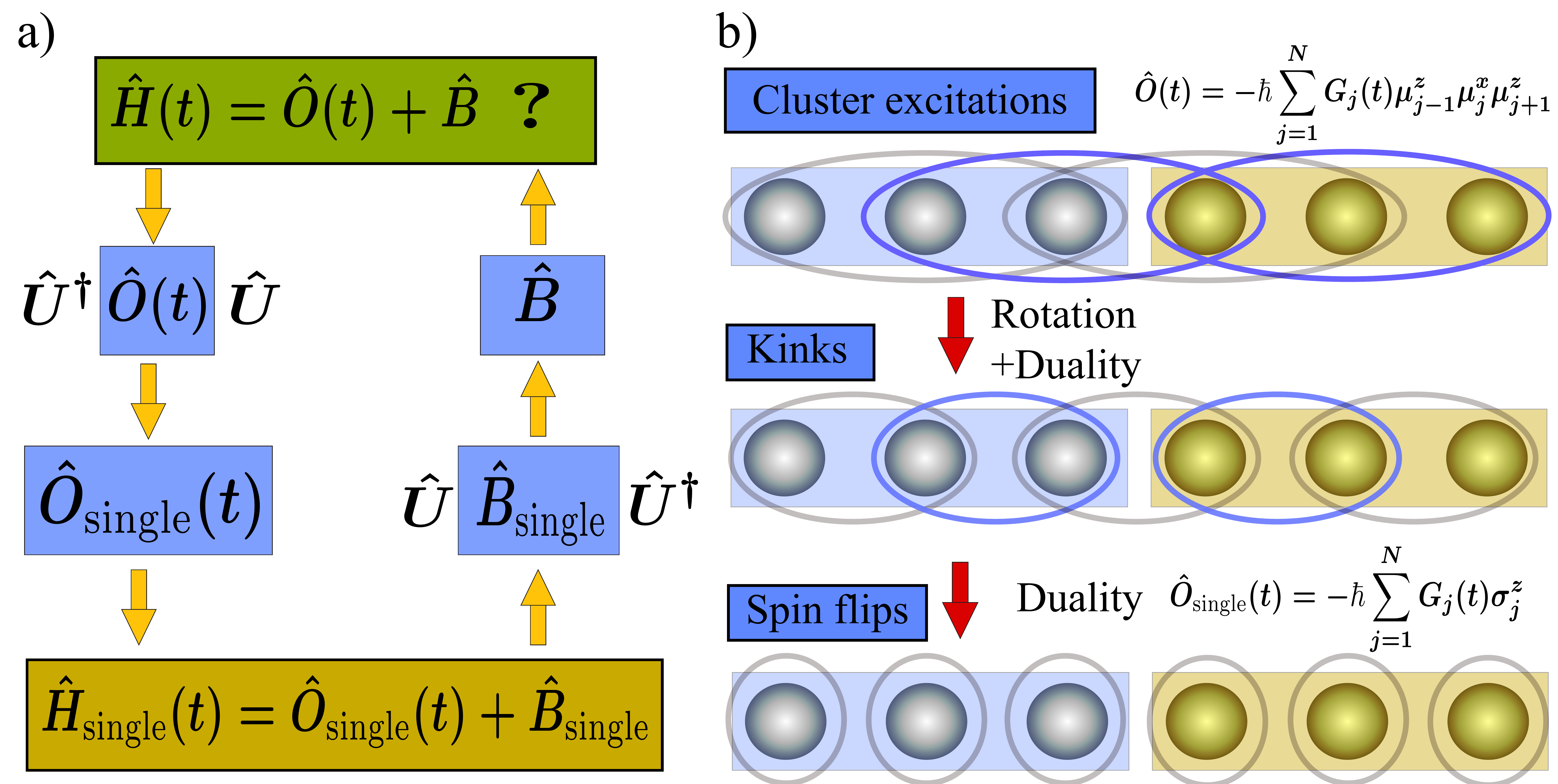}
	\caption{Topological pumping of highly correlated states. \idg{a} Depicts our scheme, which allows us to pump correlations of the operator $\hat{O}(t)$. The main challenge is to find an operator $\hat{B}$ that enables the pumping protocol. To do this, one needs to find a unitary transformation $\hat{U}$ that maps the many-body operator $\hat{O}(t)$ to a single-particle operator  $\hat{O}_\text{single}(t)$.  In the single particle picture we know what operator $\hat{B}_\text{single}$ is required in order to pump particles. By transforming the operator back, we can construct the operator $\hat{B}$. \idg{b} Shows a particular example where duality transformations enable us to develop a protocol to pump cluster states. Under duality, the cluster excitations map to kinks and finally to spin flips excitations, which can be mapped to single particles for which pumping is known.}
	\label{Overview}
\end{figure}

In this  paper, we address the problem of topological pumping of quantum correlations. The cornerstone of our approach is to map the highly-correlated states that we want to pump to uncorrelated ones, for which topological pumping protocols are well known [see Fig.~\ref{Overview}~a)]. Thus, the topological pumping of correlated states can be obtained by a suitable inverse mapping. However, in general, finding the right mapping is a non-trivial problem, since it has to reduce the correlations of the states, while preserving the topological properties of the system. Here, we show  that dualities can be exploited to achieve this goal~\cite{1941Wannier1,1941Wannier2,Kogut1979,1980Savit,1996STROMINGER,  2011Nussinov,  2016SEIBERG,2016Sacramento,2017Murugan,2017Motrunich}. 
While the duality can change the entanglement of the collective excitations~\cite{son2009quantum,2017Motrunich},  the topological properties of the energy bands are unaltered. 
In this way, we can extend the idea of topological pumping  to transport highly entangled excitations.  Due to its topological nature, the proposed quantum pumping is protected against disorder and some type of interactions.

To illustrate our approach, we show how to perform topological pumping of  cluster and kink excitations~\cite{2011Sachdev,son2011quantum}, which are related to spin flips by duality, as depicted in Fig.~\ref{Overview}~b). 
We are now able to explore the pumping dynamics of these entangled excitations for the first time. We show that bipartite entanglement is  dramatically affected by duality and pumping:  spin flips become highly entangled at the anti-crossings of the spectrum. Contrary to this, the entanglement present in kinks and cluster states is reduced or stays constant for specific bipartite divisions. One of the most appealing features of topological pumping is its robustness against disorder. In most cases, introducing interactions between excitations destroys topological transport~\cite{nakagawa2018breakdown}. However, we find that for a particular type of interaction, topological pumping is still possible and furthermore, robustness against certain types of disorder is strongly enhanced.

\begin{figure}
	\includegraphics[width=0.4\textwidth]{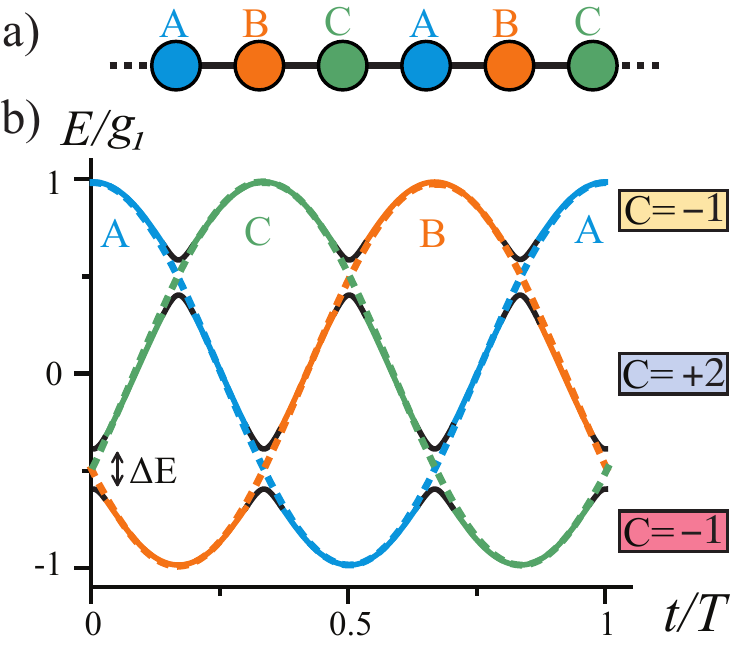}
	\caption{Topological pumping of spin excitations. a) Depicts trimers, where the sites A, B and C have energies $E_\text{A}=g_1\cos(\omega t)$, $E_\text{B}=g_1\cos(2\pi/3+\omega t)$ and $E_\text{C}=g_1\cos(4\pi/3+\omega t)$, respectively.  b) Shows the instantaneous energies of a single spin-flip excitation per trimer in the limit ${g_0\gg J}$.  The dashed lines show the local energies of the sites A (blue), B (orange) and C (green) for ${J=0}$.
The solid lines show the spectrum for ${J\ne0}$. At the crossings, the energies of two neighboring spins are degenerate. The spin-spin interaction hybridizes the excitations and lifts the degeneracy, creating an energy gap ${\Delta E\approx 2J}$. The number of anti-crossings is related to the Chern number $C$ of the respective band. For trimes,  particles are pumped by $3C$ sites per driving period.} 
	\label{Fig3}
\end{figure}

\section{Topological pumping and highly-correlated states}
In the following, we discuss our general procedure, which is schematically presented  in Fig.\ref{Overview}~a). We consider a general Hamiltonian of the form 
\begin{equation}
\hat{H}(t)=\hat{O}(t)+\hat{B}
\ ,
\end{equation}
where the operator $\hat{O}(t)$  creates many-body correlations in the system. Our goal is to pump these correlations. In principle, the task can be achieved by finding a time-independent operator $\hat{B}$ that is non-commuting with $\hat{O}(t)$. However, to find the right operator can be non-trivial, because in general, not all operators  $\hat{B}$ allow topological pumping. The operator  $\hat{B}$ must couple the eigenstates of $\hat{O}(t)$ to lift the degeneracies between them at all times $t$ during the pumping process. 
In this paper, we aim to find a unitary transformation $\hat{U}$ that allows us to map the correlated problem to a simpler one, for which the topological pumping protocol is well known.  
The unitary operator $\hat{U}$ has to be sufficiently non-local in space to un-correlate $\hat{O}(t)$.  If such unitary mapping can be obtained, then  $\hat{O}(t)$ can be transformed to a single-particle operator $\hat{O}_\text{single}(t)$. Then,  the corresponding single-particle pumping operator $\hat{B}_\text{single} $ can be found by resorting to known frameworks of topological pumping of single particles
\begin{equation}
\hat{H}_{\text{single}}(t)=\hat{O}_\text{single}(t)+\hat{B}_\text{single} 
\ .
\end{equation}
Finally, the correlation pumping operator $\hat{B}$ is obtained via the inverse transformation $\hat{B}=\hat{U}\hat{B}_\text{single}\hat{U}^{\dagger}$.
We discuss below an example of our general scheme, where $\hat{U}$ is constructed by combining duality and rotations in a spin system. In order to highlight the capability of our scheme, we show below how to perform topological pumping of cluster  and kink states.

\subsection{Mapping cluster states to spin flips}

Let us begin by considering the time-dependent operator  $\hat{O}(t)=-\hbar\sum^{N}_{j=1}G_j(t)\mu^z_{j-1}\mu^x_{j}\mu^z_{j+1}$, where ${\mu^\alpha_j}$ (${\alpha\in\{x,y,z\}}$) are Pauli matrices acting on the $j$-th site. The eigenstates of this operator are cluster-like states, which are highly-entangled objects relevant for quantum computation~\cite{raussendorf2003measurement} or quantum memories~\cite{goihl2019harnessing}.
In order to be able to pump cluster states, we need to find a suitable unitary operator $\hat{U}$. With this aim, let us consider the Kramers-Wannier duality
$\tau^z_{j}\tau^z_{j+1}= \mu^x_{j}$, 
$\tau^x_{j}
= \mu^z_{j}
\mu^z_{j+1} $, 
$ \tau^y_{j}\tau^y_{j+1}=\mu^z_{j}\mu^x_{j+1}\mu^z_{j+1}$. Here, $\tau_j^\alpha$ are the Pauli matrices after the dual transformation~\cite{1941Wannier1,1941Wannier2}. By applying the duality transformation, rotating the spins around the $x$ axis and then applying the duality again (see Fig.\ref{Overview}b), one can remove the correlations in  $\hat{O}(t)$. This allows us to map the latter to a single-particle operator  $\hat{O}_{\text{single}}(t)=-\hbar\sum^{N}_{j=1}G_j(t)\sigma^x_{j}$ in terms of new Pauli matrices $\sigma^x_{j}$. In contrast to the original operator $\hat{O}(t)$, the eigenstates  $\hat{O}_{\text{single}}(t)$ are uncorrelated states known as spin flips. Now it is quite direct to find an operator $\hat{B}_\text{single}$ enabling us to perform topological pumping of spin flips. To do this, let us consider the one-dimensional quantum Ising model in a transverse field~\cite{2011Sachdev}
\begin{equation}
\label{SpinflipIsing}
\hat{H}^{\text{flip}}(t)=\hbar\sum^{N}_{j=1}\left(- G_j(t)\sigma^x_{j}+ J \sigma^z_{j}
\sigma^z_{j+1}\right)\ , 
\end{equation}
which is a paradigmatic model in condensed matter and statistical physics. We consider a transverse field strength ${G_j(t) = g_0+g_1\cos[2\pi (j-1)b+\omega t+\phi_0]}$, which is adiabatically modulated in time. Here $\omega$ is the frequency of the drive and its period is ${T=2\pi/\omega}$. The parameters $\phi_0$ and $1/b$ determine the initial phase shift and the spatial period of the modulation, respectively. In addition,  $J$ is strength of spin-spin interaction. We assume periodic boundary conditions $\sigma^\alpha_{j}=\sigma^\alpha_{j+N}$ and explore the specific case $b=1/3$. However, our results remain valid for any rational value $b=p/q$, where $p$ and $q$ are coprimes. In the following, we will show that $\hat{B}_\text{single}=\hbar J\sum^{N}_{j=1}\sigma^z_{j}
\sigma^z_{j+1}$ is the operator that allows us to perform topological pumping of spin flips.

\subsection{Topological pumping of cluster states}

Due to the spatial modulation of the transverse field, the
Ising Hamiltonian of Eq.~\eqref{SpinflipIsing} exhibits topological features that
we can exploit to perform topological pumping. To gain some intuition about this, let us consider the weak-coupling regime $g_0\gg J$, where the collective excitations are spin-flips. 
Since the interaction between neighboring spins is small, the total number of excitations  $\hat{\mathcal{N}}=1/2\sum^{N}_{j=1}(1+\sigma^x_{j})$ is approximately conserved. Thus, the spectrum is divided into different bands associated with a fixed number of spin flips (see Fig.~\ref{Fig3}). Within a given band, the Hamiltonian~\eqref{SpinflipIsing} exhibits the same dynamics as the Aubry-Andre (AA) model~\cite{aubry80,kraus2012}, which is a toy model for topological pumping~\cite{thouless1983,Thouless1984}. The current of transported particles is intimately related to the Chern numbers  $C$ of the Harper-Hofstadter model~\cite{1976Hofstadter}. In appendices~\ref{AppendixA}~and~\ref{AppendixB} we describe these models in detail. During the pumping process, the Chern number $C$ determines the number of lattice sites that a single particle is transported per cycle~\cite{xiao2010berry}. In other words, the mean position changes as  $x(T)-x(0)=1/2\sum^{N}_{j=1} j(\langle\sigma^x_{j}(T)\rangle-\langle\sigma^x_{j}(0)\rangle)=C$, 
where $C$ is the Chern number associated to the relevant band,  $x(t)=\langle\hat{x}(t)\rangle$, and $\hat{x}=1/2\sum^{N}_{j=1} j(\sigma^x_{j}+1)$ is the position operator. For our choice of $G_j(t)$ with a trimer structure, the particle is transported $3C$ sites per driving period.

\begin{figure}
	\includegraphics[width=0.45\textwidth]{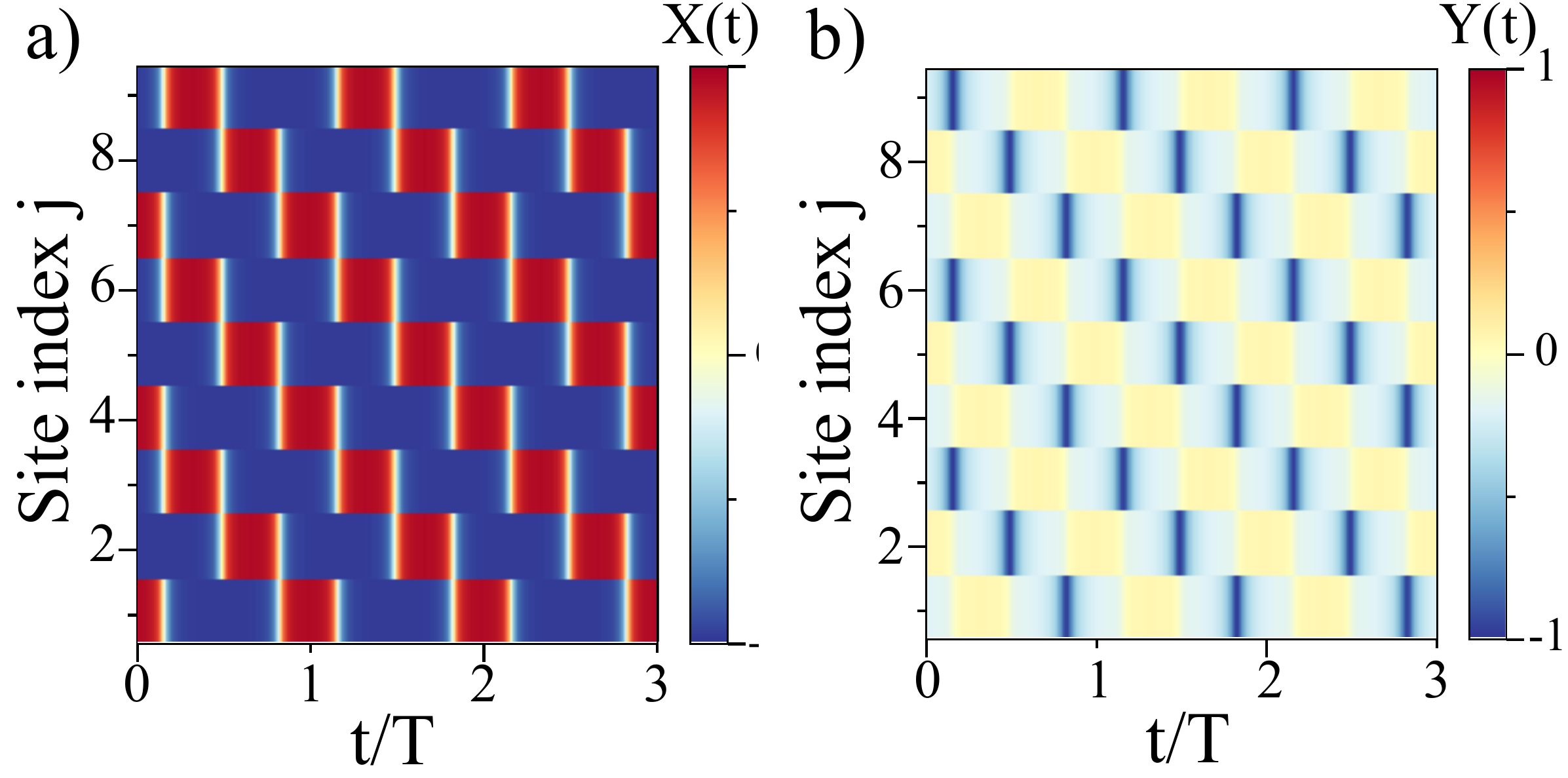}
	\caption{ a) Evolution of the expectation value of the driving operator of the Hamiltonian. For spin-flip (Eq. \eqref{SpinflipIsing}) $X_j(t)=\mean{\sigma^x_j}$, for kink (Eq. \eqref{KinkIsing}) $X_j(t)=\mean{\tau^z_j\tau^z_{j+1}}$ and for cluster-Ising model (Eq. \eqref{DualClusterHamiltonian}) $X_j(t)=\mean{\mu^z_j\mu^x_{j+1}\mu^z_{j+2}}$. b) Expectation value of the non-driven operator of the model Hamiltonians. For spin flips $Y_j(t)=\mean{\sigma^z_j\sigma^z_{j+1}}$, for kink model $Y_j(t)=\mean{\tau^x_j}$ and for cluster-Ising model $Y_j(t)=\mean{\mu^z_j\mu^z_{j+1}}$. 	The initial state is the eigenstate with one excitation per trimer in the lowest band with Chern number ${C=-1}$. Increasing or decreasing the number of trimers in the chain does not substantially affect the dynamics. Parameters for all graphs are ${N=9}$, ${g_0=10J}$, ${g_1=3J}$, initial phase ${\phi_0=0}$, and frequency  ${\omega=0.02J}$.}
	\label{PumpingObservables9}
 \end{figure}

From our previous discussion, we conclude that the operator $\hat{B}_\text{single}=\hbar J \sum^{N}_{j=1} \sigma^z_{j}
\sigma^z_{j+1}$ allows us to perform topological pumping of single particle excitations. By applying the inverse of the above transformations, we can construct the operator $\hat{B}=\hbar J \sum^{N}_{j=1} \mu^z_{j}
\mu^z_{j+1}$ that allows us to pump cluster states. This gives rise the cluster-Ising Hamiltonian~\cite{skrovseth2009phase,doherty2009identifying,smacchia2011statistical,son2012topological} 
\begin{align}
\label{DualClusterHamiltonian}
\hat{H}^{\text{cluster}}(t)&=\hbar\sum^{N}_{j=1}\left(-G_j(t)\mu^z_{j-1}\mu^x_{j}\mu^z_{j+1}+J \mu^z_{j}
\mu^z_{j+1}\right)
\ .
\end{align}
By repeatedly applying spin rotations  and duality transformations,  there is a plethora of complex many-body spin operators that can be reduced to single particle operators, as we show in appendix~\ref{AppendixC}. These models host  excitations with a wide range of topological phases~\cite{nie2017scaling}, for which we can now construct a pumping protocol.  As an example, by using the duality, we can find a Hamiltonian to pump kinks, which are eigenstates of the operator $\hat{O}(t)=-\hbar\sum^{N}_{j=1}G_j(t)\tau^z_{j} \tau^z_{j+1}$. In fact, as the Hamiltonian
\begin{align}
\label{KinkIsing}
\hat{H}^{\text{kink}}(t)&=\hbar\sum^{N}_{j=1}\left(-G_j(t)\tau^z_{j} \tau^z_{j+1}+J \tau^x_{j}\right)
\ .
\end{align}
is dual to the Ising Hamiltonian~\eqref{SpinflipIsing}, the natural choice of the operator that allows us to pump kinks is $\hat{B}=\hbar J \sum^{N}_{j=1} \tau^x_{j}$. 
During topological pumping, the change of the mean position is related to the Chern number. In the case of Hamiltonian~\eqref{DualClusterHamiltonian}, we can show that correlations can be pumped (see appendix~\ref{AppendixB}), as follows
\begin{equation}
         \label{eq;CenterMassClusters}
         x(T)-x(0)=\frac{1}{2}\sum^{N}_{j=1} j(\langle\mu^z_{j}\mu^x_{j+1}\mu^z_{j+2}(T)\rangle-\langle\mu^z_{j}\mu^x_{j+1}\mu^z_{j+2}(0)\rangle)=C
         \ .
\end{equation}
To illustrate this, we depict the dynamics of topological pumping of quantum correlations in Fig.~\ref{PumpingObservables9}~a).

 \begin{figure}
	\includegraphics[width=0.48\textwidth]{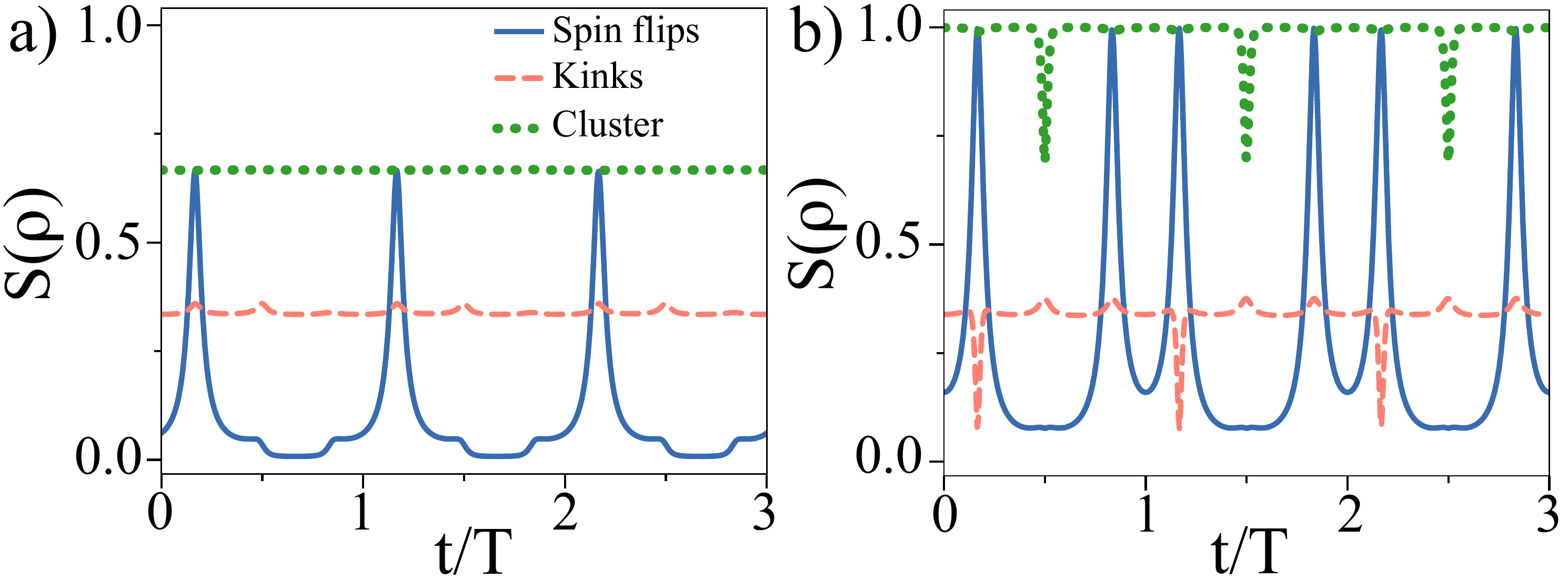}
	\caption{ Evolution of von-Neumann entropy for the dual models with two types of partitions: a) first trimer and b) first site of each trimer is traced out. The degree of entanglement as measured by the von-Neumann entropy is normalized relative to the maximal possible entanglement. The parameters are the same as in Fig.~\ref{PumpingObservables9}.
	}
	\label{PumpingEntanglement9}
 \end{figure}

\begin{figure}[htbp]
	\centering
	\subfigimg[width=0.38\textwidth]{}{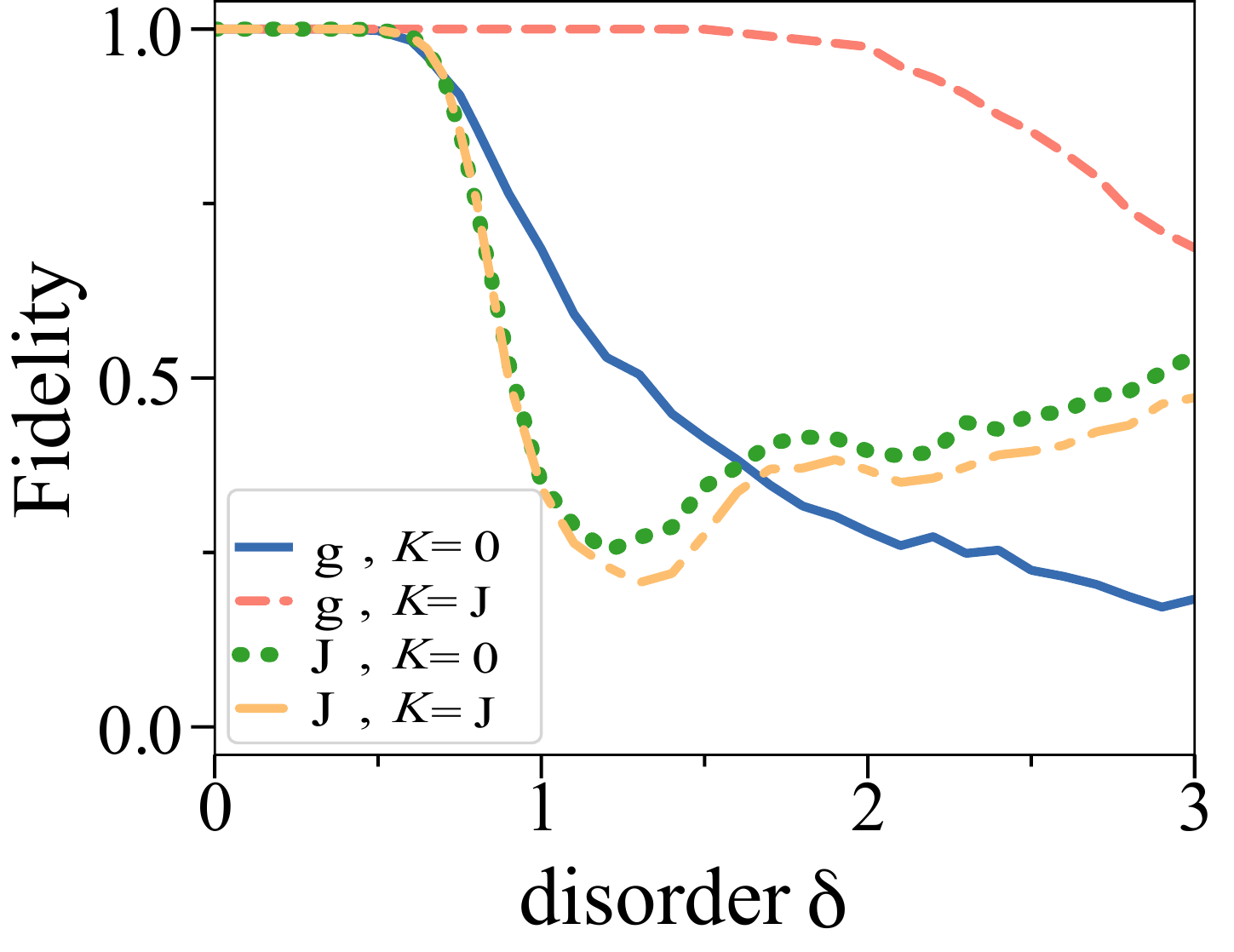}\hfill
	\caption{ Fidelity of the state ${F=\abs{\braket{\Psi(0))}{\Psi(3T)}}^2}$ after 3 pump cycles for random disorder strength $\delta$  applied either to $G_j(t)$ or $J$ of the models. Disorder is sampled randomly between $[-\delta,\delta]$.
	Without interaction $K$, fidelity starts to decay when disorder is on the order of the energy gap ${\delta_c\approx0.7J}$, independent of which variable is disordered. With interaction $K$, pumping is more robust for disorder in $G_j(t)$, however unchanged for $J$. 
	Parameters are ${N=9}$, ${g_1=3J}$, initial phase ${\phi_0=0}$, and angular frequency  ${\omega=0.02J}$.  }
	\label{DisorderNew}
\end{figure}
\section{Entanglement dynamics and topological pumping}

\begin{table*}[htbp]
	\centering
	\begin{center}
		\begin{tabular}{c|c|c} 
			&away from anti-crossing&at anti-crossing\\ \hline
			spin-flip& $\ket{010}\otimes\dots\otimes\ket{010}_x$& $\frac{\sqrt{3}}{\sqrt{2L}}\left[(\ket{01}-\ket{10})\ket{0}\otimes\dots\otimes(\ket{01}-\ket{10})\ket{0}_x\right]$ \\
			\hline
			kink&$\frac{1}{\sqrt{2}}(\ket{100}\otimes\dots\otimes\ket{100}+\ket{011}\otimes\dots\otimes\ket{011}_z)$ &$\frac{\sqrt{3}}{\sqrt{4L}}[\ket{1}(\ket{0}-\ket{1})\ket{0}\otimes\dots\otimes\ket{1}(\ket{0}-\ket{1})\ket{0}_z+$\\ 
			&&$\ket{0}(\ket{0}-\ket{1})\ket{1}\otimes\dots\otimes\ket{0}(\ket{0}-\ket{1})\ket{1}_z]$ \\ \hline
			cluster-Ising&$\prod_{j=0}^{N/3} \mu^z_{3j+2}\prod_{i=1}^N\text{c}\mu^z_{i,i+1}\ket{000}\otimes\dots\otimes\ket{000}_x$&
			$\prod_{j=0}^{N/3} \frac{1}{\sqrt{2}}(\mu^z_{3j+2}-\mu^z_{3j+1})\prod_{i=1}^N\text{c}\mu^z_{i,i+1}\ket{000}\otimes\dots\otimes\ket{000}_x$
		\end{tabular}
		\caption{Eigenstates with one excitation per trimer in the pumping process of the spin-flip, kink and cluster-Ising models. We show the analytic states at and away from the anti-crossing in the limit ${g_0\gg J}$. States are given in terms of the eigenstates of the $z$-basis $\ket{0}$ and $\ket{1}$ for kinks and in the $x$ basis for spin-flip and cluster-Ising model. For the cluster-Ising model we define the ground state of cluster model as ${\prod_{i=1}^N\text{c}\mu^z_{i,i+1}\ket{000}_x}$, with $\text{c}\mu^z_{i,i+1}$ being the control phase gate acting on site $i$, ${i+1}$. 
		}\label{tab:states}
	\end{center}
\end{table*}
Kinks and clusters are highly entangled states, whereas spin flips are very close to product states. 
A natural question that arises is: what is the dynamics of entanglement during topological pumping and how this depends on the character of the excitations? 
To answer this question we divide the spin chain into two subsystems A and B, perform a partial trace over A $\hat{\rho}_\text{B}=\Tr_\text{A}(\hat{\rho})$ and calculate the von-Neumann entropy $S=-\Tr_\text{B}(\hat{\rho}_\text{B}\log\hat{\rho}_\text{B})$~\cite{smacchia2011statistical}. Here $\hat{\rho}$ is the density matrix of of the total system. We consider two types of partitions: In partition 1 [Fig.~\ref{PumpingEntanglement9}a)], A is the first trimer, and B the rest. In partition 2 [Fig.~\ref{PumpingEntanglement9}b)], A is the first site of each trimer, and B the rest. The system is translational invariant for each trimer. Thus, these two partitions highlight the local and extended entanglement properties of the spins.
We find that the entanglement exhibits dramatic changes during the pumping process. In particular, the entanglement  of the spin flips increases at the anti-crossings. Then, the spin-flips are in a superposition state of being at two sites, which could be measured in experiments~\cite{bloch2016, takahashi2016}.
In contrast, the entangled kink and cluster states loose entanglement at the anti-crossings for the case 2 partition, and have nearly unchanged entanglement for case 1. The difference arises because partition 2 extends over all trimers, whereas partition 1 is measuring only a single trimer, highlighting the non-local structure of the quantum state. These properties can also be seen in the structure of the quantum states. 
The spin-flip, kink and cluster-Ising Hamiltonians are related by duality transformation, which changes the wavefunction and entanglement profoundly.  We can calculate the states for a single excitation per trimer exactly in the limit ${g_0\gg J}$ in Table \ref{tab:states} at and away from anti-crossings.  
For our chosen values (${g_0=10J}$, ${g_1=3J}$), we find an overlap of the analytic states with the numeric results of more than 95\% both at and away from the anti-crossing. The overlap decreases with increasing $g_0$ or $J$ and is nearly unaffected by system size as we checked up to ${N=9}$.

\section{The effect of disorder and interactions}

\begin{figure}[h]
	\centering
	\subfigimg[width=0.38\textwidth]{}{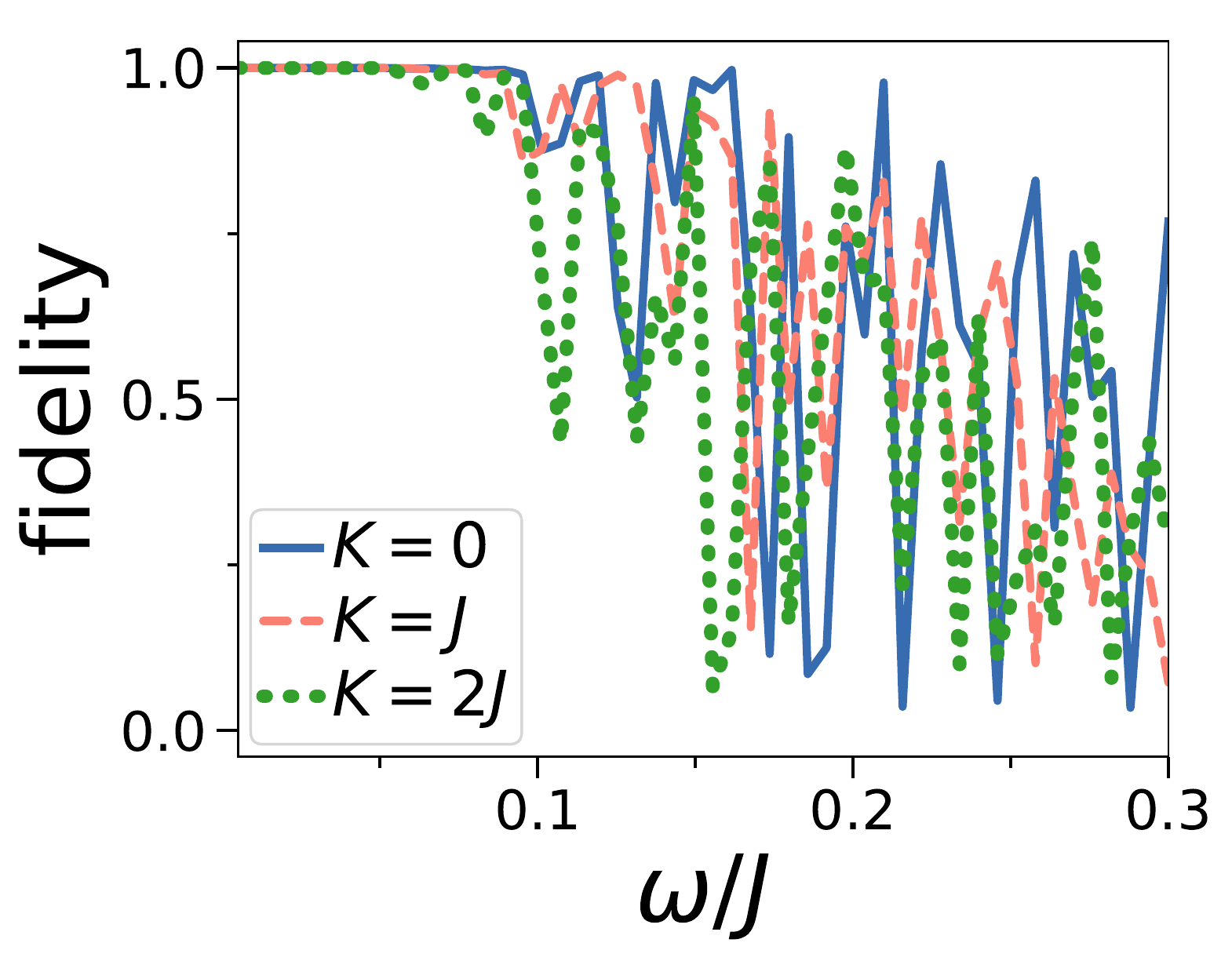}\hfill
	\caption{Fidelity of the state ${F=\abs{\braket{\Psi(0))}{\Psi(9T)}}^2}$ after 9 pump cycles for varying driving frequency $\omega$ and different interaction strength $K$. For $\omega<0.1J$, pumping is stable. For higher frequencies, the fidelity starts to oscillate.
	Parameters are ${N=9}$, ${g_1=3J}$ and initial phase ${\phi_0=0}$.  }
	\label{frequency}
\end{figure}

Topological pumping is generally considered with non-interacting particles. It is known that interactions can break topological pumping in some situations~\cite{nakagawa2018breakdown}, while  it is still possible in specific cases~\cite{tangpanitanon2016topological,haug2018topological}. While some intuition about this question is present for the single-particle case, how does interaction affect pumping of non-local quantum correlations? Our scheme can shed light on this question, as the duality transformation can relate different types of interactions.
We find that the lowest band  with one excitation per trimer (Chern number ${C=-1}$) can be pumped even when adding the interaction term ${\hat{H}^\text{flip}_\text{int}=\hbar K\sum_j\sigma^x_j\sigma^x_{j+1}}$ to the spin-flip Hamiltonian $\hat{H}^\text{flip}$. However, interaction destroys pumping in  the other bands. The pumping of that specific band persists as the single excitations of each trimer are highly localized and separated by a large distance of 3 sites. Thus, they couple only weakly to each other and the energy gap is nearly unchanged compared to the non-interacting case. This reasoning is valid for any model that can be reduced to spin-flips via the transformation $\hat{U}$. The corresponding transformed interaction terms are ${\hat{H}^\text{kink}_\text{int}=\hbar K\sum_j\sigma^z_j\sigma^z_{j+2}}$ for kink Hamiltonian Eq.\eqref{KinkIsing} and ${\hat{H}^\text{cluster}_\text{int}=\hbar K\sum_j\sigma^z_j\sigma^y_{j+1}\sigma^y_{j+2}\sigma^z_{j+3}}$ for cluster-Ising Eq.\eqref{DualClusterHamiltonian}.
Interaction has a profound impact on the robustness to disorder. We implement random spatial disorder $\Delta_j$ with strength $\delta$ in either variable ${G_j'(t)=G_j(t)+\Delta_j}$ or ${J'=J+\Delta_j}$ of our Hamiltonians, where $\Delta_j$ is randomly sampled between $[-\delta,\delta]$.
In Fig.~\ref{DisorderNew} we show the fidelity ${F=\abs{\braket{\Psi(0)}{\Psi(3T)}}^2}$ of the pumped state after 3 pump cycles. We observe that without interaction ${K=0}$, the fidelity is reduced above a critical disorder, independent of whether disorder is applied to $G_j(t)$ or $J$. However, with interaction ${K=J}$, pumping is much more stable for disorder applied to $G_j(t)$. This  effect cannot be attributed to a change in the energy gap, as it is nearly unchanged with interaction $K$. However, we observe that the system is more robust against disorder when the interaction operator commutes with the part of the Hamiltonian that is disordered, i.e., for spin-flip model $\hat{H}^\text{flip}_\text{int}$ commutes with $\sigma^x_i\sigma^x_{i+1}$, but not with $\sigma^z_i\sigma^z_{i+1}$. We conjecture that the increased stability arises when the interaction Hamiltonian acts as a stabilizer on the pumped state, and leads to a renormalization of disorder due to interaction $g$ \cite{yue1994conduction}. 
These findings on the stability against disorder found for spin-flips can be immediately applied to pumping of kinks and cluster-Ising via the mapping scheme. In Fig.\ref{frequency}, we illustrate the frequency dependence of the pumping. We measure the overlap between the initial state and the pumped state after 9 pumping cycles. 
Fidelity of the pumping is unity below a critical driving frequency. Interaction decreases the critical driving frequency slightly. Above this frequency, we observe strong oscillations in the pumping fidelity with frequency.

\section{Conclusions}

In conclusion, we have proposed a scheme to perform topological pumping of quantum correlations. Our approach is based on a unitary transformation that maps correlated states to uncorrelated ones, where known protocols for topological pumping can be used. To illustrate our general idea, we exploit spin dualities to realize topological pumping of cluster states, which are states relevant for quantum computation~\cite{raussendorf2003measurement} or quantum memories~\cite{goihl2019harnessing}. We also show that the pumping process can reduce the bipartite entanglement for kinks and cluster excitations, but enhances it for single spin-flips. Recent experiments have realized topological pumping of particles~\cite{bloch2016, takahashi2016}, without considering the entanglement. It would be of considerable interest to revisit these experiments to measure the entanglement dynamics.
The spin-flip and kink models we considered are experimentally realizable in current quantum simulators~\cite{2014Nori,2017Noh,2016Hartmann} that are available in several platforms such as superconducting qubits~\cite{roushan17, Martinis2018,Fan2018}, trapped ions~\cite{Monroe2016}, Rydberg~\cite{Weimer2011} and cold atoms~\cite{Bloch2015,Gross2016,Amico_NJP,dumke2016roadmap}. Cluster states have been realized in photonic systems~\cite{chen2007experimental} as well cluster Hamiltonians could be realized with cold atoms~\cite{pachos2004three}. Our approach could be applied to find pumping schemes for other types of quantum correlations and to transport them, while being protected against disorder. Robustness against specific types of disorder can be enhanced by certain interaction terms, which can be determined by using our method. It would be interesting to investigate the implication of our results for error correction in spin chains~\cite{Bowen2019} and the role of other type of dualities~\cite{Almheiri2015} on topological pumping.

 \appendix

\section{The Harper-Hofstadter model and topological pumping in the Aubry-Andre model}
\label{AppendixA}
\begin{figure}[h]
\vspace{0.4cm}
	\includegraphics[width=0.45\textwidth]{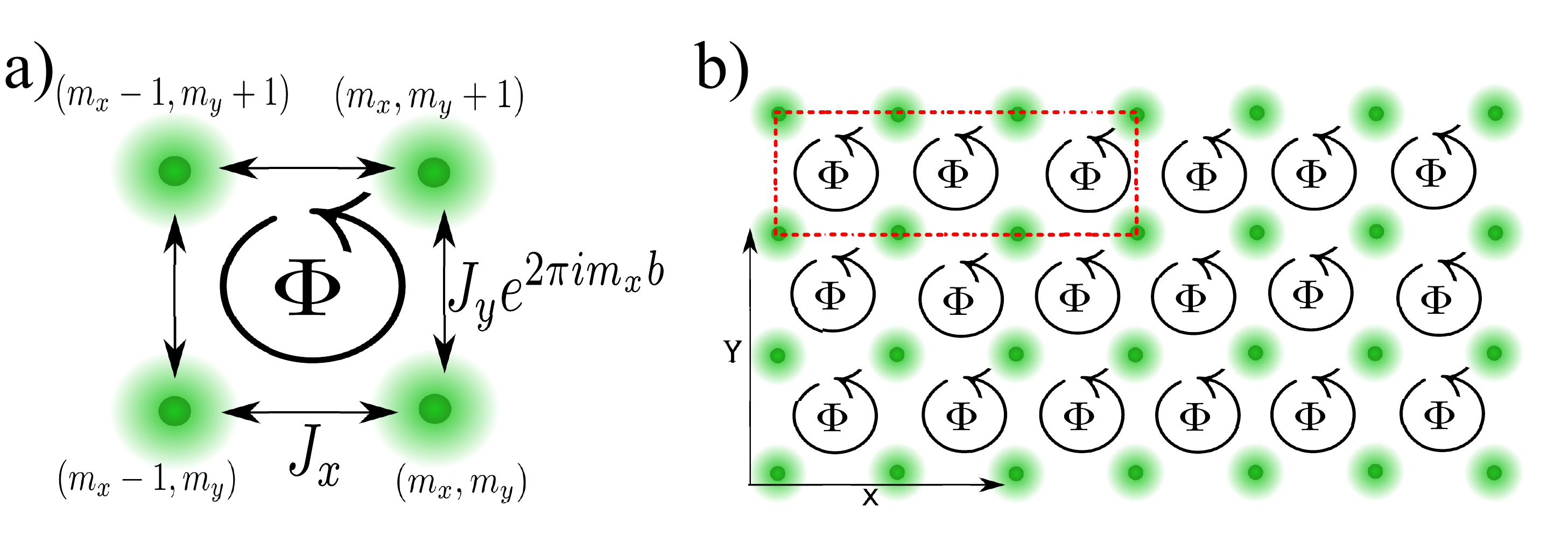}
    \caption{The Harper-Hofstadter model of electrons in a two-dimensional lattice. a) Depicts the flux through a unit cell and b) the lattice described by the Harper-Hofstadter model. }
    \label{Fig1}
\end{figure}

The Harper-Hofstadter model describes a Bloch electron moving in a two-dimensional lattice under the effect of a magnetic field
\begin{align}
\label{eq:2DHarper}
    \hat{H}_{\text{HH}}&=-\hbar J_x\sum_{m_x,m_y}(f^{\dagger}_{m_x,m_y} f_{m_x+1,m_y}-f_{m_x,m_y} f^{\dagger}_{m_x+1,m_y}+\text{h.c})
    \\&\nonumber
    -\hbar J_y \sum_{m_x,m_y}(e^{2i\pi m_x b}f^{\dagger}_{m_x,m_y}f_{m_x,m_y+1}-e^{-2i\pi m_x b}f_{m_x,m_y}f^{\dagger}_{m_x,m_y+1})
  \ .
 \end{align}
 Here $f_{m_x,m_y}$ and $f^{\dagger}_{m_x,m_y}$ are fermionic operators and $J_x$ and $J_y$ are hoppings strengths along x and y directions, respectively.  When the electron completes a loop in the lattice, it acquires a phase $\phi$ proportional to the magnetic flux through the loop, as it is depicted in figure~\ref{Fig1}. Within a unit cell, the flux is denoted as $\phi=2\pi b$, where $b$ is a real number. The physics of the Harper model is extremely rich and it is closely related to the integer quantum Hall effect and topological pumping. The reason of the non-trivial features of the model is the character of the real number $b$. For example, if it is a rational $b=p/q$ with $p,q$ integers, the magnetic flux through a unit cell is commensurable with the flux quanta. Figure~\ref{Fig1} (b) shows the case $b=1/3$, where the system is decomposed in terms of a sublattice.  In this case, one can transform the Hamiltonian to quasimomentum representation, as follows
\begin{equation}
\label{eq:HarperQuasimomenum}
       \hat{H}_{\text{HH}}=\sum_{k_x,k_y}\Psi^{\dagger}_{k_x,k_y} \hat{H}_{k_x,k_y}^{\text{HH}}\Psi^{\dagger}_{k_x,k_y} 
       \ ,
\end{equation}
where $\Psi^{\dagger}_{k_x,k_y}=(F^{\dagger}_{A,k_x,k_y},F^{\dagger}_{B,k_x,k_y},F^{\dagger}_{C,k_x,k_y})$. The operators $F^{\dagger}_{n,k_x,k_y}$ with $n\in\{A,B,C\}$ are the Fourier transform of the fermionic operators in real space. Due to our choice $b=1/3$, the Hamiltonian $\hat{H}_{k_x,k_y}^{\text{HH}} $ is a $3\times 3$ matrix and
the system exhibits three bands that we denote by A, B and C respectively. For a given band, $n\in\{A,B,C\}$, one can calculate the associated Chern number
\begin{equation}
         \label{eq:2DChernNumber}
                C_n=\frac{1}{2\pi}\int_{\text{FBZ}}\Omega_n(k_x,k_y)\text{d}k_x\text{d}k_y\, ,
\end{equation}
with the Berry curvature 
\begin{align}
         \label{eq:2DChernCurv}
\Omega_n(k_x,k_y)&=i(\braket{\partial_{k_x} u_n(k_x,k_y)}{\partial_{k_y} u_n(k_x,k_y)}
\\&\nonumber
-\braket{\partial_{k_y} u_n(k_x,k_y)}{\partial_{k_x} u_n(k,t)})
\ ,
\end{align}
 and $u_n(k_x,k_y)$ being the eigenstate of the $n$-th band of the model in Fourier space.

An important step to establish the intimate relation between the Harper-Hofstadter model and topological pumping, is to reduce the dimension of the system from a two-dimensional lattice to a set of decoupled one-dimensional lattices. To achieve this,
we assume periodic boundary conditions along the $y$-direction, one can transform the $y$ direction in Eq.~\eqref{eq:2DHarper} into momentum representation, as follows
\begin{align}
\label{eq:1DHarper}
\hat{H}_{HH}&=-2\hbar J_y\sum_{m_x,k_y}\cos(2\pi b m_x+k_y)(f^{\dagger}_{m_x,k_y}f_{m_x,k_y}-f_{m_x,k_y}f^{\dagger}_{m_x,k_y})
\\&\nonumber
-\hbar J_x\sum_{m_x,k_y}(f^{\dagger}_{m_x,k_y}f_{m_x+1,k_y}-f_{m_x,m_y}f^{\dagger}_{m_x+1,k_y})
=\sum_{k_y}\hat{H}(k_y)
\ .
 \end{align}
 
 This Hamiltonian $\hat{H}(k_y)$ is very important for our work, because now we can modulate the quasimomentum $k_y=\omega t+\phi_0$ adiabatically in time to perform topological pumping. After this procedure, we obtain the Aubry-Andre model
 
  \begin{align}
 \label{eq:HarperDimRed}
         \hat{H}_{AA}(t)&=-2\hbar J_y\sum_{j}\cos(2\pi b j+\omega t+\phi_0)[f^{\dagger}_{j}f_{j}+\text{H.c}]
         \\&\nonumber
         -\hbar J_x\sum_{j}(f^{\dagger}_{j}f_{j+1}+\text{H.c})
         \ .
 \end{align}
 Note that as we are working with a one dimensional system, we consider a index $j$ to label the lattice sites. Similarly to the 2D Harper-Hofstadter model Eq.\eqref{eq:2DHarper}, for our choice $b=1/3$, the Hamiltonian Eq.\eqref{eq:HarperDimRed} exhibits 3 bands. This can be derived by Fourier transforming the annihilation and creation operators along the lattice index. Its bands have non-trivial topological Chern numbers, which are defined as
\begin{equation}
C_n=\frac{1}{2\pi}\int_0^{2\pi}\int_0^T\Omega_n(k,t)\text{d}t\text{d}k\, ,
\end{equation}
with the Berry curvature $\Omega_n(k,t)=i(\braket{\partial_t u_n(k,t)}{\partial_k u_n(k,t)}-\braket{\partial_k u_n(k,t)}{\partial_t u_n(k,t)})$ and $u_n(k,t)$ being the eigenstate of the $n$-th band of the model in Fourier space\cite{xiao2010berry,bloch2016}. We neglected the band index $n$ for simplicity in the main text. The anomalous velocity $\dot{x}=\Omega_n\omega$ gives the speed with which the particles moves during the pumping process. To understand the meaning of this quantity, let us consider an initial state $\ket{\psi(0)}$, representing a particle localized at a site $j_0$ along the lattice. Depending on the position $j_0$, the mean energy of the particle should be in one of the bands A,B or C of the system. Now we can define the operator $\hat{x}=\sum_{j} j f^{\dagger}_{j}f_{j}$ in such a way that $x(t)=\bra{\psi(t)}\hat{x}\ket{\psi(t)}$. After a single period $T$ of the time modulation, the position of the particle changes proportionally to the Chern number, as $x(T)-x(0)=C_n$, where n is the band index and $x(0)$ is proportional to $j_0$  \cite{aidelsburger2015measuring} (see also Fig.3 in main text).

\section{Mapping the Ising model to the Aubry-Andre model in the limit of weak spin interactions}
\label{AppendixB}
The excitations above the paramagnetic ground state of the spin flip Hamiltonian $\hat{H}^{\text{flip}}(t)$ in Eq.(3) of the main text are spin-flips.  
In general, in the one-dimensional quantum Ising model the number of spin flips $\hat{\mathcal{N}}^{\text{flips}}=1/2\sum^{N}_{j=1}(1-\sigma^x_{j})$ is not conserved. However, if the coupling $J$ between the spins in the $z$-direction is weak, $\hat{\mathcal{N}}^{\text{flips}}$ is approximately conserved, and we can obtain an effective Hamiltonian that preserves the number of spin flips, even under an adiabatic modulation of the onsite energies ${G_{j}(t) = g_0+g_1\cos[2\pi (j-1)b+\omega t+\phi_0]}$. In order to do this, we go to a rotating frame with the unitary operator $\hat{V}(t)=\exp\left(-\mathrm{i} g_0 t\sum^{N}_{j=1}\sigma^x_{j}\right)$, where the Hamiltonian is given by $\hat{H}_{R}^{\text{flip}}(t) =\hat{V}^{\dagger}(t)(\hat{H}^{\text{flip}}(t)-\mathrm{i}\hbar \partial_t)\hat{V}(t)$. After neglecting fast oscillating terms in $\hat{H}_{R}^{\text{flip}}(t)$ with frequencies proportional to $g_0$ and after going back to the laboratory frame, we obtain an effective Hamiltonian of the form
\begin{align}
\label{eq:EffectiveDualSpinHamiltonian}
\hat{H}^{\text{flip}}(t)&\approx \hbar \sum^{N}_{j=1}\left[-G_{j}(t)\sigma^x_{j}+J( \sigma^z_{j}
\sigma^z_{j+1}+\sigma^y_{j}\sigma^y_{j+1})\right]
\ ,
\end{align}
which is valid in the limit $g_0\gg J$. Under the rotating wave approximation (RWA), one just keeps terms in the Hamiltonian that preserve the number of spin flips. In this case, one can use the Jordan-Wigner transformation

\begin{equation}
 \sigma^z_l = -f^{\dagger}_l e^{\mathrm{i}\hat{\Phi}_l}-f_l e^{-\mathrm{i}\hat{\Phi}_l},\,\  \sigma^y_l = -\mathrm{i}f^{\dagger}_l e^{\mathrm{i}\hat{\Phi}_l}+\mathrm{i}f_l e^{-\mathrm{i}\hat{\Phi}_l}, \,\   \sigma^x_l =2f^{\dagger}_l f_l-1 \ ,
\end{equation}
with $\hat{\Phi}_l=\sum_{j < l}f^{\dagger}_j f_j$ to map the Eq.\eqref{eq:EffectiveDualSpinHamiltonian} to the Aubry-Andre model discussed above 
\begin{equation}
\label{eq:HarperHamiltonian}
\hat{H}^{\text{flip}}(t)\approx2\hbar\sum^{N}_{j=1}\left[-G_{j}(t)f^{\dagger}_{j}f_{j}+J(f^{\dagger}_{j}f_{j+1}+\text{h.c.})\right] \, ,
\end{equation}
where $f^{\dagger}_{j}$ ($f_{j}$) are fermionic creation (annihilation) operators. With the definition of $G_j(t)$, this model maps  to the Aubry-Andre model of Eq.\eqref{eq:HarperDimRed}, which has been widely studied for topological pumping.

In the previous section, we discussed the relation between the Chern number and the change in the mean position $x$ of particle. As we are working with spin flips here, we can define a single spin flip initially localized at a given site $j_0$. The position operator $\hat{x}=\sum_{j} j f^{\dagger}_{j}f_{j}$ of the spin flip is given by $\hat{x}=1/2\sum^{N}_{j=1} j(\sigma^x_{j}+1)$ and its mean value reads $x(t)=\bra{\psi(t)}\hat{x}\ket{\psi(t)}$, where $\ket{\psi(t)}$ represent the state during the pumping process.  During topological pumping, the change of the position is related to the Chern number. In our case, we obtain the relation  $x(T)-x(0)=1/2\sum^{N}_{j=1} j(\langle\sigma^x_{j}(T)\rangle-\langle\sigma^x_{j}(0)\rangle)=C_n$, 
where $C_n$ is the Chern number. Now if we take into account the duality, we can show that a similar relation is satisfied for cluster like states
\begin{equation}
         \label{eq;CenterMassClusters}
         x(T)-x(0)=\frac{1}{2}\sum^{N}_{j=1} j(\langle\mu^z_{j}\mu^x_{j+1}\mu^z_{j+2}(T)\rangle-\langle\mu^z_{j}\mu^x_{j+1}\mu^z_{j+2}(0)\rangle)=C_n
         \ .
\end{equation}
In this case, however, the center of mass of the excitation, i.e., its mean position, is given by a three-point correlation function. A similar expression can be obtained for kinks, where
the center of mass is obtained by calculating a two-point correlation function.

\section{Higher order dualities}
\label{AppendixC}
Our procedure can also reduce more complex models to single particles. For example, let us consider the manybody operator
\begin{align*}
\hat{O}(t)&=-\hbar(-1)^r\sum^{N}_{j=1}G_{j}(t)\tilde{\mu}^z_{j-1}\left[\prod_{m=0}^{r}\tilde{\mu}^x_{m+j}\right]\tilde{\mu}^z_{j+1+r}
\ ,
\end{align*}
where $r$ is a positive integer.
By repeatedly applying $\pi$ rotations around the $x$-axis of the Pauli matrices and then applying the duality, it can be subsequently reduced to spin-flips $\hat{O}_{\text{single}}(t)=-\hbar\sum_jG_{j}(t)\sigma^x_j$. After this reduction, it is quite simple to find the operator $\hat{B}_{\text{single}}=\hbar J\sum_j\sigma^z_j\sigma^z_{j+1}$ that allows us to perform topological pumping of spin flips. Now we can apply the inverse transformation and obtain the operator $\hat{B}=\hbar(-1)^r J\sum^{N}_{j=1} \tilde{\mu}^z_{j-1}\left[\prod_{m=0}^{r-1}\tilde{\mu}^x_{m+j}\right]\tilde{\mu}^z_{j+r}$. By considering the aforementioned operators, we can construct the Hamiltonian
\begin{align*}
\hat{H}(t)&=-\hbar(-1)^r\sum^{N}_{j=1}G_{j}(t)\tilde{\mu}^z_{j-1}\left[\prod_{m=0}^{r}\tilde{\mu}^x_{m+j}\right]\tilde{\mu}^z_{j+1+r}+
\\&\nonumber
\hbar(-1)^r J\sum^{N}_{j=1} \tilde{\mu}^z_{j-1}\left[\prod_{m=0}^{r-1}\tilde{\mu}^x_{m+j}\right]\tilde{\mu}^z_{j+r}
\ ,\numberthis\label{eq:DualHigherHamiltonian}
\end{align*}
which allows us to pump highly correlated states. As a direct consequence, the $(r+2)$-point correlation function $\Delta_{j-1,j,\ldots,j+1+r}(t)=\langle\tilde{\mu}^z_{j-1}\left[\prod_{m=0}^{r}\tilde{\mu}^x_{m+j}\right]\tilde{\mu}^z_{j+1+r}(t)\rangle$ satisfies
\begin{equation}
         \label{eq;CenterMassClusters}
         x(T)-x(0)=\frac{1}{2}\sum^{N}_{j=1} j[\Delta_{j-1,j,\ldots,j+1+r}(T)-\Delta_{j-1,j,\ldots,j+1+r}(0)]=C_n
         \ .
\end{equation}
This equation establishes a link between quantum correlations and the Chern number, which is a topological quantity.

\begin{acknowledgments}
We thank D. G. Angelakis, M. Estarellas, M. Hanks and H. Price for valuable discussions. 
 We thank National Research Foundation Singapore, the
Ministry of Education Singapore Academic Research Fund Tier 2 (Grant No.
MOE2015-T2-1-101), and the Japanese QLEAP program for support.
The Grenoble LANEF framework
(ANR-10-LABX-51-01) is acknowledged for its support with mutualized
infrastructure. The computational work for this article was partially performed on resources of the National Supercomputing Centre, Singapore (https://www.nscc.sg).
\end{acknowledgments}

\end{document}